\documentclass[prd,superscriptaddress,amsfonts,amssymb,amsmath,showpacs,twocolumn]{revtex4-2}
\usepackage{bm}
\usepackage{amsfonts}
\usepackage{latexsym}
\usepackage{graphicx}
\usepackage{amsmath}
\usepackage{palatino}
\usepackage{mathpazo}
\usepackage{textcomp}
\usepackage{float}
\usepackage{booktabs}
\usepackage{dcolumn}
\usepackage{booktabs}
\usepackage{multirow}
\usepackage{hyperref}
\hypersetup{colorlinks,citecolor=blue}
\usepackage{amsmath}
\usepackage{xcolor}
\usepackage{orcidlink}
\usepackage[caption=false]{subfig}
\usepackage{commath}
\def\jnl@style{\it}
\def\aaref@jnl#1{{\jnl@style#1}}

\def\aaref@jnl#1{{\jnl@style#1}}

\def\aj{\aaref@jnl{AJ}}                   
\def\apj{\aaref@jnl{ApJ}}                 
\def\apjl{\aaref@jnl{ApJ}}                
\def\apjs{\aaref@jnl{ApJS}}               
\def\apss{\aaref@jnl{Ap\&SS}}             
\def\aap{\aaref@jnl{A\&A}}                
\def\aapr{\aaref@jnl{A\&A~Rev.}}          
\def\aaps{\aaref@jnl{A\&AS}}              
\def\mnras{\aaref@jnl{Mon.~Not.~Roy.~Astron.~Soc.}}             
\def\prd{\aaref@jnl{Phys.~Rev.~D}}        
\def\prc{\aaref@jnl{Phys.~Rev.~C}}  
\def\prl{\aaref@jnl{Phys.~Rev.~Lett.}}    
\def\qjras{\aaref@jnl{QJRAS}}             
\def\skytel{\aaref@jnl{S\&T}}             
\def\ssr{\aaref@jnl{Space~Sci.~Rev.}}     
\def\zap{\aaref@jnl{ZAp}}                 
\def\nat{\aaref@jnl{Nature}}              
\def\aplett{\aaref@jnl{Astrophys.~Lett.}} 
\def\apspr{\aaref@jnl{Astrophys.~Space~Phys.~Res.}} 
\def\physrep{\aaref@jnl{Phys.~Rep.}}      
\def\physscr{\aaref@jnl{Phys.~Scr}}       
\def\commat{\aaref@jnl{Comm.~Math.~Phys.}}              
\def\science{\aaref@jnl{Science}}               
\def\cqg{\aaref@jnl{Classical Quant.~Grav.}}            
\def\jpcs{\aaref@jnl{JPCS}}                                     
\def\ijmpd{\aaref@jnl{Int.~J.~Mod.~Phys.~D}}                    
\def\grg{\aaref@jnl{Gen.~Relat.~Gravit.}}               
\def\rpp{\aaref@jnl{Rep.~Prog.~Phys.}}          
\def\npa{\aaref@jnl{Nucl.~Phys.~A}}        
\def\lrr{\aaref@jnl{Living Rev.~Rel.}}                   
\def\jcap{\aaref@jnl{J.~Cosmology Astropart.~Phys.}}    
\def\rmp{\aaref@jnl{Rev.~Mod.~Phys.}}   
\def\epjc{\aaref@jnl{Eur.~Phys.~J.~C}}

\allowdisplaybreaks[1]

\begin{document}

\color{black}       

\title{A New $Om(z)$ Diagnostic of Dark Energy in General Relativity Theory}

\author{N. Myrzakulov\orcidlink{0000-0001-8691-9939}}
\email[Email: ]{nmyrzakulov@gmail.com}
\affiliation{L. N. Gumilyov Eurasian National University, Astana 010008,
Kazakhstan.}
\affiliation{Ratbay Myrzakulov Eurasian International Centre for Theoretical
Physics, Astana 010009, Kazakhstan.}

\author{M. Koussour\orcidlink{0000-0002-4188-0572}}
\email[Email: ]{pr.mouhssine@gmail.com}
\affiliation{Quantum Physics and Magnetism Team, LPMC, Faculty of Science Ben
M'sik,\\
Casablanca Hassan II University,
Morocco.} 

\author{Dhruba Jyoti Gogoi \orcidlink{0000-0002-4776-8506}}
\email[Email: ]{moloydhruba@yahoo.in}
\affiliation{Department of Physics, Dibrugarh University,
Dibrugarh 786004, Assam, India.}


\date{\today}

\begin{abstract}
In this paper, we propose a new parametrization of dark energy based on the $Om(z)$ diagnostic tool behavior. For this purpose, we investigate a functional form of the $Om(z)$ that predicts the popular dark energy dynamical models, namely phantom and quintessence. We also found the famous cosmological constant for specified values of the model's parameters. We employed the Markov Chain Monte Carlo approach to constrain the cosmological model using Hubble, Pantheon samples, and BAO datasets. Finally, we used observational constraints to investigate the characteristics of dark energy evolution and compare our findings to cosmological predictions.
\end{abstract}

\maketitle

\section{Introduction}
\label{sec1}

The General Theory of Relativity (GR) by Albert Einstein is a magnificent achievement that has been validated by many years of experimental testing \cite{Will}. Despite the efficacy of GR in characterizing the Universe and the solar system, it is widely agreed that GR, along with the cosmological constant ($\Lambda$), is just an exceptionally excellent estimate valid within the current range of experimental observations. Lately, Modified Gravity Theories (MGT) have received a lot of attention in the hopes of finding observationally compatible alternatives to GR. This is owing to new observational findings such as Type Ia supernovae (SNeIa) \cite{Riess, Perlmutter}, Baryon Acoustic Oscillations (BAO) \cite{D.J., W.J.}, Cosmic Microwave Background (CMB) \cite{R.R., Z.Y.}, Large Scale Structure (LSS) \cite{T.Koivisto, S.F.}, and the Planck collaborations \cite{Planck2020}, indicating the existence of two unexplained components that may influence the evolution of the Universe. In this regard, measurements have resulted in the addition of new exotic 
fluids such as Dark Energy (DE) of large negative pressure, which leads to the accelerated expansion of the Universe, and dark matter, which is the cause of the formation of galaxies clusters, inside the standard model of cosmology. On the other hand, the unclear nature of these constituents can also be regarded as the possibility of GR collapse on an enormous scale. The $\Lambda$CDM (Lambda Cold Dark Matter) model is probably the most simple cosmological model that includes these two dark constituents. A cosmological constant is added to the standard Einstein-Hilbert action in this scenario with the Equation of State (EoS) $\omega_{\Lambda}=-1$. So, if one assigns the cosmological constant to vacuum energy, it suffers from a "fine-tuning" problem, which relates to the difference between the observed and theoretically expected values of the $\Lambda$ \cite{dalal/2001, weinberg/1989}. This issue has fueled impulses to look for alternate DE models outside the $\Lambda$CDM model.

There are two major ways to work with such issues: one involves different components inside the GR action and then studies the possible impacts that may develop, such as scalar fields, vector fields, or other matter field types \cite{M.C., A.Y., T.Chiba, C.Arm., Carroll, Fujii}. In addition, altering the background theory and analyzing subsequent equations of motion is another option for discovering novel characteristics that may not be consistent with astronomical data, such as $f(R)$ gravity, $f(T)$ gravity, and $f(Q)$ gravity \cite{R1, g01, R2, R3, T1, T2, T3, T4, T5, QQ1, QQ2,Kou1,Kou2, Kou3, Kou4}. Recently, several studies have been done in different modified theories of gravity in different aspects \cite{Kou5, dj1,dj2,dj3,dj4,dj5}.

However, several studies have attempted to investigate the evolution of the Universe without relying on any certain cosmological model. Such approaches are sometimes referred to as model-independent ways study of cosmological models or cosmological parametrization \cite{Cunha, Mortsell}. To find the exact solutions of Einstein field equations, this approach is generally based on the assumption of parametrization of geometrical parameters (such as the Hubble parameter $H$, deceleration parameter $q$, jerk parameter $j$, and so on) or physical parameters (such as the energy density $\rho$, pressure $p$, EoS parameter $\omega$, and so on). The approach has no effect on the background theory and clearly provides solutions to the Einstein field equations. It also has the benefit of reconstructing the cosmic evolution of the Universe and explaining some of its features. Furthermore, this approach gives the easiest way to theoretically overcome several of the standard model's issues, including the initial singularity problem, the cosmological constant problem, and the late-time acceleration scenario. In the literature, there are numerous ways of DE parameterization, such as: see \cite{H1, H2, H3, H4} for the Hubble parameter, see \cite{q1, q2, q3, q4, q5, q6} for the deceleration parameter, see \cite{j1, j2, j3} for the jerk parameter, and see \cite{EoS1, EoS2, EoS3, EoS4, EoS5, EoS6, EoS7, EoS8} for the EoS parameter. Ref. \cite{Pacif} summarizes a large number of different parameterization methods.

Following the approach, Sahni et al. \cite{Sahni1} introduced a successful diagnostic called $Om(z)$, which is responsive to the EoS of DE and so offers a null test of the $\Lambda$CDM model and has been intensively researched in numerous publications \cite{Sahni2, Ding, Zheng}. When the value of this diagnostic tool remains constant for all redshift values, DE takes the form of a $\Lambda$, but varying $Om(z)$ corresponds to various dynamical DE scenarios. However, the slope of $Om(z)$ can differentiate between two types of DE models: a positive slope suggests phantom phase ($\omega_{DE}<-1$), whereas a negative slope shows quintessence ($\omega_{DE}>-1$) \cite{Sahni1}. Several previous studies have used reconstructed $Om(z)$ with the combination of Gaussian processes and observations such as Hubble datasets, SNeIa datasets, and BAO datasets to undertake compatibility tests of the $\Lambda$CDM model \cite{Seikel, Yahia}. So, it is important to employ some parametrization to analyze the $Om(z)$ diagnostic in a cosmological model-independent context. This method has both benefits and drawbacks. One advantage is that it is not affected by the Universe's matter and energy content. One shortcoming of this formulation is that it does not describe the source of the accelerated expansion \cite{Jesus1}.

In this paper, we investigate a new parametrization of the $Om(z)$ diagnostic and discuss the cosmic evolution in the framework of GR. The $Om(z)$ diagnostic functional form is constructed such that it predicts the popular DE dynamical models, namely phantom and quintessence. The behavior of the $Om(z)$ diagnostic is determined by the model parameters that were constrained by the observational data. Here, we consider 31 data points of the Hubble expansion observations performed using the differential age approach \cite{Sharov/2018} and BAO data that include six points \cite{BAO1}. Scolnic et al. \cite{Scolnic/2018} published recently Pantheon, a huge SNe Ia datasets with 1048 points across the redshift range $0.01<z<2.26$. The Hubble, Pantheon samples, and BAO datasets with the Markov Chain Monte Carlo (MCMC) approach are used in our study to constrain the cosmological model.

The following is how this work is organized: In Sec. \ref{sec2}, we describe briefly the newly suggested $Om(z)$ diagnostic parametrization, then apply it to a homogeneous and isotropic Universe in the framework of GR theory. In Sec. \ref{sec3}, we use the MCMC approach to constrain the model parameters using Hubble datasets, Pantheon datasets, BAO datasets, and combinations such as Hubble+Pantheon datasets and Hubble+Pantheon+BAO datasets. Sec. \ref{sec4} starts with a review of observational constraints and a discussion of findings. Lastly, Sec. \ref{sec5} concludes with some final remarks.

\section{Cosmological model}

\label{sec2}

In this section, we present the essential cosmological scenario equations
for our model. The Friedmann-Lema\^{\i}tre-Robertson-Walker (FLRW) model is
the fundamental mathematical framework of cosmology, describing a
homogeneous and isotropic Universe in which everything is the same in all
directions and at all points. The metric for a spatially flat Universe is
expressed as, 
\begin{equation}
ds^{2}=dt^{2}-a^{2}(t)[dr^{2}+r^{2}(d{\theta }^{2}+sin^{2}\theta d{\phi }%
^{2})],  \label{eq:2.2}
\end{equation}%
where, $r$, $\theta $, and $\phi $ are the spatial coordinates, $t$ is the time coordinate, and $a(t)$ is the scale factor that represents the expansion of the
Universe. For the purpose of simplicity, we have fixed the scale factor to 1 currently. However, it is important to note that the scale factor itself is not observable. What is observable is the ratio of the scale factor at any given time to its value at some reference time, often taken to be the present time. For convenience, we have chosen to set the value of the scale factor at the present time $a_0$ to 1. This choice is equivalent to referring to the ratio of the scale factor at any given time to its value at the present time $a/a_0$.

In addition, the energy-momentum tensor of a perfect fluid (with no
viscosity) defines the fluid's energy density and pressure. It is presented
by%
\begin{equation}
T_{\mu \nu }=\left( p+\rho \right) u_{\mu }u_{\nu }-pg_{\mu \nu },
\label{EMT}
\end{equation}%
where $\rho $ is the energy density, $p$ is the isotropic pressure of the
Universe, $u^{\mu }$ is the fluid's 4-velocity, and $g_{\mu \nu }$ is the
metric tensor. The indices $\mu $\ and $\nu $ vary between $0$ and $3$. If
the fluid is at repose $u^{\mu }=\left\{ 1,\overrightarrow{0}\right\} $,
then $T_{00}=\rho $ and $T_{ij}=-pg_{ij}$.

The Einstein field equations for GR are given by%
\begin{equation}
R_{_{\mu \nu }}-\frac{1}{2}g_{_{\mu \nu }}R=\kappa T_{\mu \nu },
\label{EFEs}
\end{equation}%
where $\kappa =8\pi G=1$, $R_{_{\mu \nu }}$ is the Ricci curvature tensor,
and $R$ is the scalar curvature.

Using Eqs. (\ref{eq:2.2})-(\ref{EFEs}), the Einstein field equations for a
spatially flat FLRW Universe can be expressed as, 
\begin{equation}
3H^{2}=\rho  \label{fe1}
\end{equation}%
\begin{equation}
2{\dot{H}}+3H^{2}=-p  \label{fe2}
\end{equation}%
where $H=\frac{\dot{a}}{a}$ is the Hubble parameter which is a measure of
the Universe's current rate of expansion, and a dot denotes differentiation
with respect to cosmic time $t$. In the previous equation, $\rho $ and $p$
indicate the energy density and pressure of the Universe, respectively.
Also, Eqs. (\ref{fe1}) and (\ref{fe2}) are known as Friedmann equations. The
first Friedmann equation connects the Universe's expansion rate ($H$) to its
energy density, and the second Friedmann equation connects the acceleration
of the expansion rate to the pressure.

Now, to characterize the cosmic history and the possible transition to an
accelerated period, we use the total equation of state (EoS) parameter $\omega $%
, given as,%
\begin{equation}
\omega =\frac{p}{\rho }
\end{equation}

Using Eqs. (\ref{fe1}) and (\ref{fe2}), the EoS parameter is expressed as,%
\begin{equation}
\omega =-\frac{2{\dot{H}}+3H^{2}}{3H^{2}}=-1-\frac{2{\dot{H}}}{3H^{2}}
\end{equation}

The $Om(z)$ diagnostic, an intriguing null test of DE, was proposed in
\cite{Sahni1}. The beauty of this concept comes in its theoretical
structure, which is formed from the Hubble parameter $H(z)$, a quantity that
can be estimated from observations of various astronomical phenomena, such as SNeIa and BAO. This approach distinguishes
between the cosmological constant and dynamical models of DE. If the value
of $Om(z)$ remains constant at any redshift, DE takes the form of a
cosmological constant, but varying $Om(z)$ corresponds to various dynamical
DE scenarios. Nevertheless, the slope of $Om(z)$ can differentiate between
two sorts of DE models: a positive slope suggests phantom phase ($\omega_{DE} <-1$%
), whereas a negative slope shows quintessence ($\omega_{DE} >-1$). Several
previous research has undertaken consistency checks of the $\Lambda $CDM
model utilizing reconstructed $Om(z)$ based on the preceding conclusions
\cite{Seikel, Yahia, Shahalam, Pasqua}. Motivated by the physical evidence of the $Om(z)$ slope and
the above discussion, we propose a parametrization of $Om(z)$ written in
terms of redshift $z$ as,%
\begin{equation}
Om\left( z\right) =\alpha \left( 1+z\right) ^{n}  \label{model}
\end{equation}

Here, $\alpha $ and $n$ are the two parameters of the model. The above
formula clearly shows that $\Lambda $CDM is entirely recovered when $\alpha
=\Omega _{m}^{0}$ and $n=0$. The behavior of $Om\left( z\right) $ can be
divided into three periods based on the value of parameter $n$: quintessence
(negative slope) for $n<0$, phantom (positive slope) for $n>0$, and lastly
the cosmological constant (constant slope) for $n=0$ (please see Tab. \ref{tab1}). Also, one of the
advantages of the $Om(z)$ parametrization is that it exhibits a finite value
at $z=0$ (present). The introduction of the parameter $n$ in the above
parametrization provides a novel cosmological-model-independent method of
discriminating between a greater range of cosmological solutions with
varying EoS ($\omega_{DE} <-1$, $\omega_{DE} >-1$ and $\omega_{DE} =-1$).

\begin{table}[h]
\centering 
\begin{tabular}{cccc}
\hline\hline
$n$ & $Om(z)$ slope & $\omega_{DE}$ & Model \\ \hline
$n=0$ & Constant & $\omega_{DE}=-1$ & Flat $\Lambda$CDM \\ 
$n<0$ & Negative & $\omega_{DE}>-1$ & Quintessence \\ 
$n>0$ & Positive & $\omega_{DE}<-1$ & Phantom \\ \hline\hline
\end{tabular}%
\caption{Aspects of the $Om(z)$ diagnostic with relation to the value of $n$.}
\label{tab1}
\end{table}

The dimensionless Hubble parameter can be expressed in terms of the $Om(z)$
diagnostic as,%
\begin{equation}
E^{2}\left( z\right) =Om\left( z\right) \left[ \left( 1+z\right) ^{3}-1%
\right] +1,  \label{DH}
\end{equation}%
where $E\left( z\right) =\frac{H\left( z\right) }{H_{0}}$, and $H_{0}$ is
the present value of the Hubble parameter.

Now, by using Eqs. (\ref{model}) and (\ref{DH}) we have,%
\begin{equation}
E^{2}\left( z\right) =\alpha \left[ \left( 1+z\right) ^{3}-1\right] \left(
1+z\right) ^{n}+1.  \label{Ez}
\end{equation}

The redshift $z$ is connected to the scale factor $a\left( t\right) $ by $%
a\left( t\right) =\left( 1+z\right) ^{-1}$. Since $z$ is connected to the
scale factor $a\left( t\right) $, it is necessary to quantify cosmological
parameters such as the energy density, pressure, EoS in terms of $z$ to
investigate the history of the Universe in more detail. Thus, the derivative
of the Hubble parameter with respect to cosmic time is expressed as,%
\begin{equation}
\overset{.}{H}=\frac{dH}{dt}=-\left( 1+z\right) H\left( z\right) \frac{%
dH\left( z\right) }{dz}.  \label{Hd}
\end{equation}

From (\ref{DH}), Eq. (\ref{Hd}) becomes,%
\begin{equation}
\overset{.}{H}=-\frac{\alpha H_{0}^{2}}{2}\left[ 3+(3+n)z(3+z\left(
3+z\right) )\right] (1+z)^{n}.  \label{Hdd}
\end{equation}

Using Eqs. (\ref{fe1}), (\ref{fe2}), (\ref{Ez}) and (\ref{Hd}), the energy
density $\rho $ and pressure $p$ can be expressed in terms of redshift as,%
\begin{equation}
\rho \left( z\right) =3H_{0}^{2}\left\{ \alpha \left[ (1+z)^{3}-1\right]
(1+z)^{n}+1\right\} ,
\end{equation}%
and%
\begin{equation}
p\left( z\right) =H_{0}^{2}\left\{ -3+\alpha \left[ 3+nz\left( 3+z\left(
3+z\right) \right) \right] (1+z)^{n}\right\} .
\end{equation}

The EoS parameter in terms of redshift $z$ for the physical model is derived
as,%
\begin{equation}
\omega \left( z\right) =\frac{-3+\alpha \left[ 3+nz\left( 3+z\left(
3+z\right) \right) \right] (1+z)^{n}}{3+3\alpha \left[ (1+z)^{3}-1\right]
(1+z)^{n}}.  \label{EoS}
\end{equation}

Moreover, the deceleration parameter $q$, a significant cosmological
quantity, is written as,

\begin{equation}
q=-1-\frac{\overset{.}{H}}{H^{2}}=\frac{1}{2}\left( 1+3\omega \right) ,
\end{equation}%
can be derived from Eq. (\ref{EoS}) as,

\begin{equation}
q\left( z\right) =-1+\frac{\alpha \left[ \left( 3+\left( 3+n\right) z\left(
3+z\left( 3+z\right) \right) \right) \right] \left( 1+z\right) ^{n}}{%
2+2\alpha \left[ 3+z\left( 3+z\right) \right] z\left( 1+z\right) ^{n}}.
\label{qz}
\end{equation}

In next section, the possibility of a transition of the Universe's expansion
from a decelerated to an accelerated state is examined. Also, Eq. (\ref{qz})
shows that the $q\left( z\right) $ is highly dependent on the values of the
model parameters, especially $\alpha $ and $n$. In general, one can
arbitrarily choose these parameters and investigate the behavior of $q(z)$
to compare them to observational datasets. However, in this study, we first
constrain the model parameters $\alpha $ and $n$ using multiple
observational datasets such as the Hubble, Pantheon, and BAO, and then we
use the best-fit values to solve the problem.

\section{Observational data}

\label{sec3}

This section discusses the observational datasets and the statistical
analysis approach which will be employed to constrain the different
parameters of the model that were previously mentioned, followed by a
discussion of the results produced from this study. In our work, we employed
current observational datasets from Hubble, Pantheon Type Ia supernovae
(SNe Ia) samples include a number of SNe Ia data points, and baryon acoustic
oscillation (BAO) observations. To evaluate the datasets, we employ
Bayesian statistical analysis and the \textit{emcee} package in Python
language to perform a Markov chain Monte Carlo (MCMC) simulation \cite%
{Mackey/2013}.

To begin, we will look at the priors on parameters, which are shown in Tab. \ref{tab4}. In addition, to find out the findings of our MCMC study, we
employed 100 walkers and 1000 steps for all datasets. The next subsections
go into further depth on the datasets and statistical analyses.

\subsection{Hubble datasets}

The well-known cosmological principle assumes that our Universe is
homogenous and isotropic on a large scale. This is the fundamental concept
of contemporary cosmology and is the basis of the aforementioned FLRW
metric. This idea has been tested multiple times in the previous several
decades and is validated by numerous cosmological observations. In the
investigation of observational cosmology, the Hubble parameter, $H=\frac{%
\overset{.}{a}}{a}$, is used to directly analyze the Universe's expansion
scenario, where $\overset{.}{a}$ denotes the derivative of the cosmic scale
factor $a$ with respect to cosmic time $t$. As a function of redshift, the
Hubble parameter $H(z)$ can be represented as,%
\begin{equation}
H(z)=-\frac{1}{1+z}\frac{dz}{dt}.
\end{equation}

Here, $dz$ is obtained from spectroscopic surveys, and hence $dt$ provides
the model-independent value of the Hubble parameter. In principle, there
really are two well-known techniques for determining the value of the Hubble
parameter values $H(z)$ at a given redshift $z$. The first is $H(z)$
extraction from line-of-sight BAO data, while the second is the differential
age (DA) approach. In this paper, we have taken $31$ points from the DA
approach in the redshift range reported as $0.07<z<2.42$ \cite{Sharov/2018}
and tabulated in Tab. \ref{tab2} with references.

Further, we used the chi-square function to obtain the best-fit values of
the model parameters $\alpha $, $n$, and $H_{0}$ (which is equal to the
maximum likelihood analysis),%
\begin{equation}
\chi _{Hubble}^{2}=\sum_{i=1}^{31}\frac{[H_{i}^{th}(\alpha
,n,H_{0},z_{i})-H_{i}^{obs}(z_{i})]^{2}}{\sigma _{Hubble}^{2}(z_{i})},
\end{equation}%
where $H_{i}^{th}$ is the theoretical value of the Hubble parameter, $%
H_{i}^{obs}$ denotes the observed value, and $\sigma _{Hubble}^{2}$ denotes
the standard error in the observed value of $H\left( z\right) $. By using
the aforementioned datasets, we computed the best-fit values of the model
parameters, $\alpha $, $n$ and $H_{0}$ as shown in Fig. \ref{H} with the $1-\sigma 
$ and $2-\sigma $ confidence level (CL) contour, and the numerical findings
for the Hubble are shown in Tab. \ref{tab4}. In addition, we have given the error
bar plot for the mentioned Hubble datasets in Fig. \ref{Hz} along with our
resulting model compared to the $\Lambda $CDM model (with $\Omega
_{m}^{0}=0.3$, $\Omega _{\Lambda }^{0}=0.7$ and $H_{0}=69$ $%
km.s^{-1}.Mpc^{-1}$) \cite{Planck2020}. The graph illustrates that our model fits the
observational Hubble datasets well.
	
\begin{widetext}
\begin{center}
\begin{table}[h]
    \begin{tabular}{||c|c|c|c||c|c|c|c||}
    \hline
    $z$ & $H(z)$ & $\sigma_H$ & Ref. & $z$ & $H(z)$ & $\sigma_H$ & Ref. \\[0.5ex]
    \hline \hline
    $0.070$ & $69$ & $19.6$ & \cite{Stern/2010} & $0.4783$ & $80$ & $99$ & \cite{Moresco/2016}\\
    \hline
    $0.90$ & $69$ & $12$ & \cite{Simon/2005} & $0.480$ & $97$ & $62$ & \cite{Stern/2010} \\
    \hline
    $0.120$ & $68.6$ & $26.2$ & \cite{Stern/2010} & $0.593$ & $104$ & $13$ & \cite{Moresco/2012} \\
    \hline
    $0.170$ & $83$ & $8$ & \cite{Simon/2005} & $0.6797$ & $92$ & $8$ & \cite{Moresco/2012}\\
    \hline
    $0.1791$ & $75$ & $4$ & \cite{Moresco/2012} & $0.7812$ & $105$ & $12$ & \cite{Moresco/2012}\\
    \hline
    $0.1993$ & $75$ & $5$ & \cite{Moresco/2012} & $0.8754$ & $125$ & $17$ & \cite{Moresco/2012} \\
    \hline
    $0.200$ & $72.9$ & $29.6$ & \cite{Zhang/2014} & $0.880$ & $90$ & $40$ & \cite{Stern/2010} \\
    \hline
    $0.270$ & $77$ & $14$ & \cite{Simon/2005} & $0.900$ & $117$ & $23$ & \cite{Simon/2005} \\
    \hline
    $0.280$ & $88.8$ & $36.6$ & \cite{Zhang/2014} & $1.037$ & $154$ & $20$ & \cite{Moresco/2012} \\
    \hline
    $0.3519$ & $83$ & $14$ & \cite{Moresco/2012} & $1.300$ & $168$ & $17$ & \cite{Simon/2005} \\
    \hline
    $0.3802$ & $83$ & $13.5$ & \cite{Moresco/2016} & $1.363$ & $160$ & $33.6$ & \cite{Moresco/2015} \\
    \hline
    $0.400$ & $95$ & $17$ & \cite{Simon/2005} & $1.430$ & $177$ & $18$ & \cite{Simon/2005} \\
    \hline
    $0.4004$ & $77$ & $10.2$ & \cite{Moresco/2016} & $1.530$ & $140$ & $14$ & \cite{Simon/2005} \\
    \hline
    $0.4247$ & $87.1$ & $11.2$ & \cite{Moresco/2016} & $1.750$ & $202$ & $40$ & \cite{Simon/2005} \\
    \hline
    $0.4497$ & $92.8$ & $12.9$ & \cite{Moresco/2016} & $1.965$ & $186.5$ & $50.4$ & \cite{Moresco/2015}  \\
    \hline
    $0.470$ & $89$ & $34$ & \cite{Ratsimbazafy/2017} &  &  &  &  \\
    \hline
    \end{tabular}
     \caption{Hubble datasets with 31 data points.}
 \label{tab2}
    \end{table}
\end{center}

\begin{figure}[H]
\centering
\includegraphics[scale=0.75]{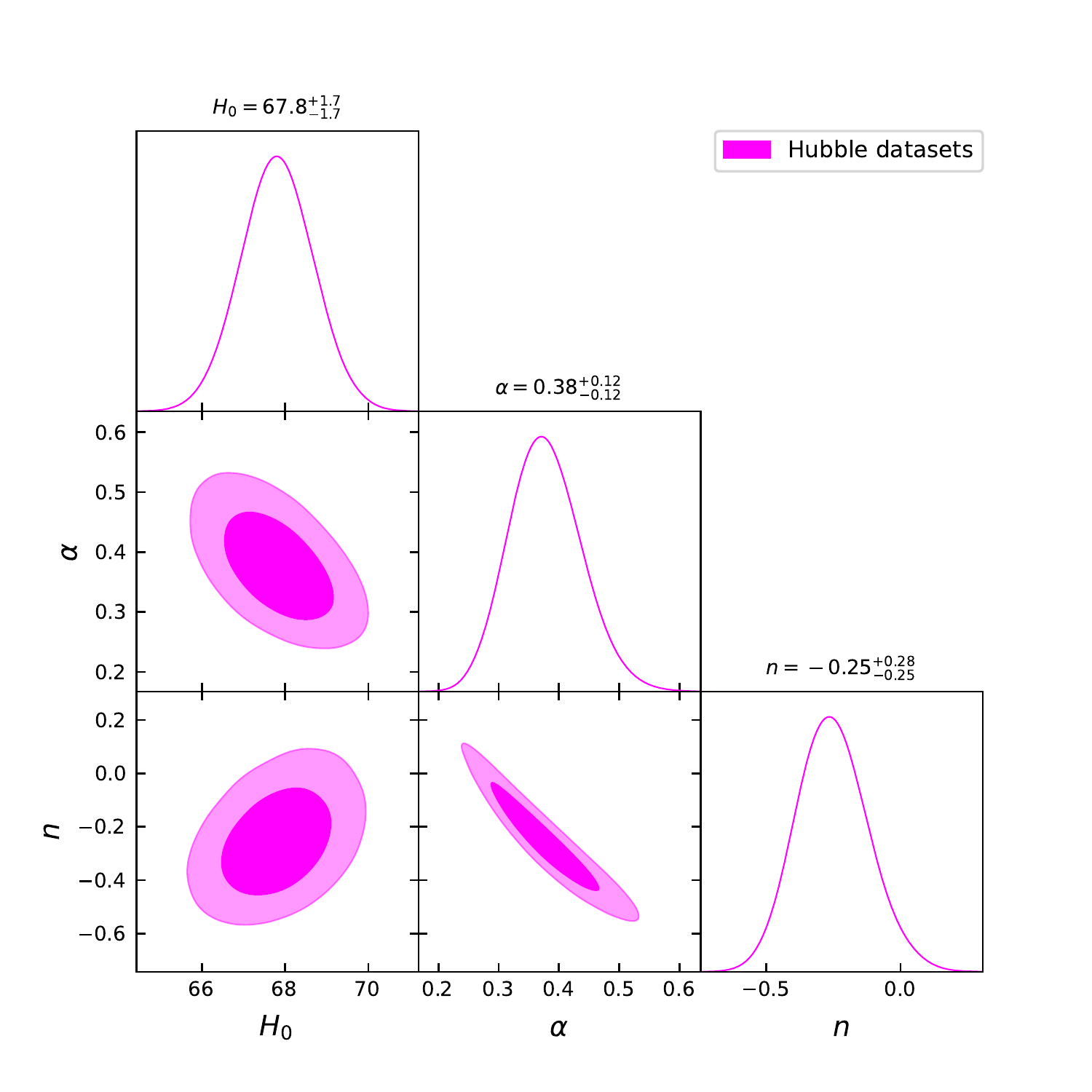}
\caption{The confidence curves at $1-\sigma$ and $2-\sigma$ and posterior distributions for the model parameters using Hubble datasets. The dark pink shaded areas represent the $1-\sigma$ confidence level (CL), while the light pink shaded areas represent the $2-\sigma$ CL. The parameter constraint values are also presented at the $1-\sigma$ CL.}
\label{H}
\end{figure}	

\begin{figure}[H]
\centering
\includegraphics[scale=0.55]{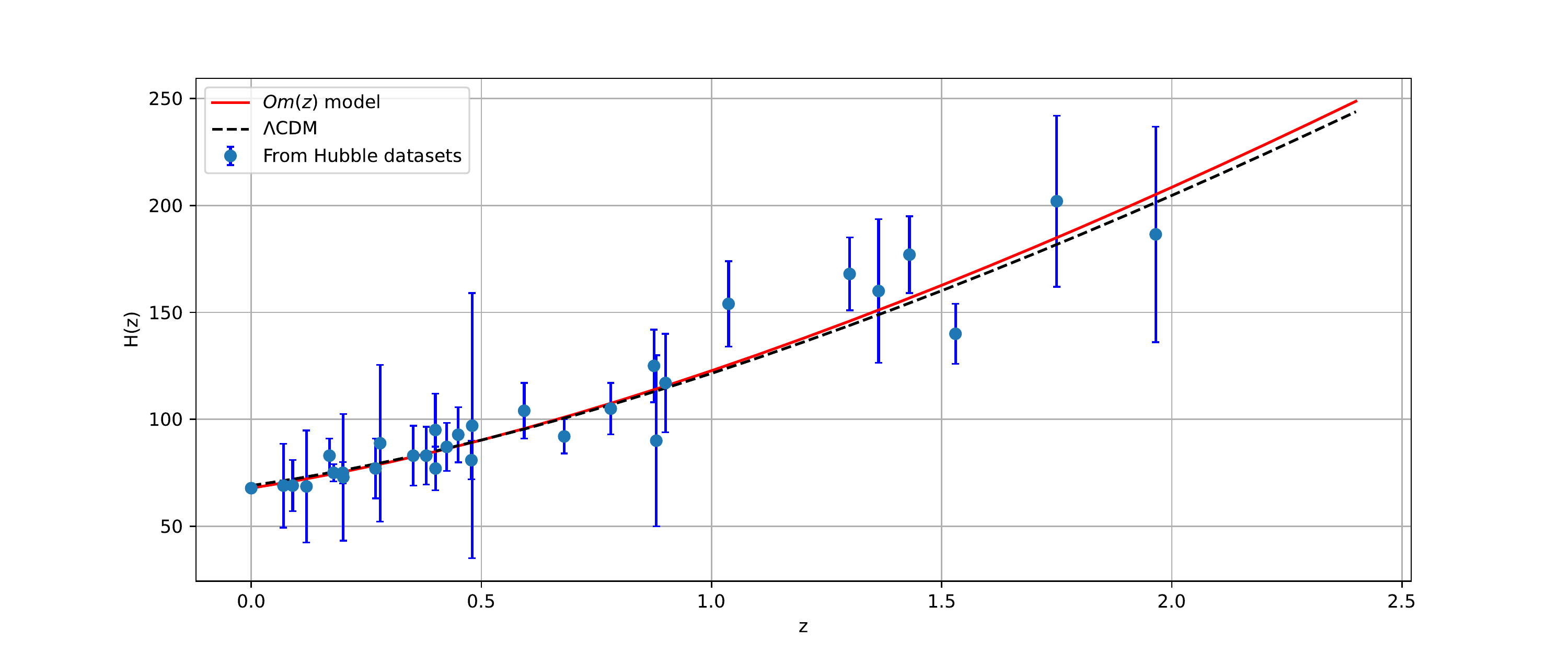}
\caption{The variation of $H(z)$ vs. $z$. The blue dots represent error bars, the red line represents our model's curve, and the black dashed line represents the $\Lambda$CDM model.}
\label{Hz}
\end{figure}	

\end{widetext}

\subsection{Pantheon datasets}

Observational research on SNe from the golden sample of 50 points of Type Ia
revealed that our Universe is expanding at a faster rate. As a result,
investigations on larger and larger samples of SNe datasets have risen
during the last 2 decades. The most recent sample of SNe Ia datasets,
consisting of 1048 data points, was just released. In this work, we used the
Pantheon datasets [83], which contain 1048 samples of spectroscopically
validated SNe Ia spanning the redshift range $0.01<z<2.26$ \cite%
{Scolnic/2018}, which combines the SNe Legacy Survey (SNLS), the Sloan
Digital Sky Survey (SDSS), the Hubble Space Telescope (HST) survey, the
Panoramic Survey Telescope, and the Rapid Response System (Pan-STARRS1).
These data points provide an estimate of the distance modulus $\mu
_{i}^{obs} $ in the redshift range $0<z_{i}\leq 1.41$. In this paper, we
compare the theoretical value $\mu ^{th}$\ with the measured value $\mu
_{i}^{obs}$ of the distance modulus to estimate our model parameters of the
produced model.

The theoretical distance modulus $\mu ^{th}$ is defined as follows:%
\begin{equation}
\mu ^{th}=\mu ^{th}\left( D_{L}\right) =m-M=5\log \left( D_{L}\right) ,
\end{equation}%
where $m$\ and $M$ indicates apparent and absolute magnitudes of a standard
candle respectively.

The luminosity distance $D_{L}(z)$ given by,%
\begin{equation}
D_{L}(z)=c(1+z)\int_{0}^{z}\frac{dz^{^{\prime }}}{H(z^{^{\prime }})}
\end{equation}

Thus, the chi-square function for the Pantheon datasets is defined as, 
\begin{equation}
\chi _{Pan}^{2}=\sum_{i,j=1}^{1048}\Delta \mu _{i}\left(
C_{Pan}^{-1}\right) _{ij}\Delta \mu _{j}.  \label{4b}
\end{equation}

Here $C_{Pan}$ is the covariance matrix \cite{Scolnic/2018}, and $%
\Delta \mu _{i}=\mu ^{th}(z_{i},\alpha ,n,H_{0})-\mu _{i}^{obs}$
is the difference between the observed distance modulus value obtained from
cosmic data and its theoretical values created from the model using the
parameter space $\alpha $, $n$, and $H_{0}$. By minimizing $\chi
_{Hubble}^{2}+\chi _{Pan}^{2}$, the constraints of the model parameters, $\alpha $%
, $n$ and $H_{0}$ from the combination
Hubble+Pantheon datasets are shown in Fig. \ref{H+SN} and numerical findings
presented in Tab. \ref{tab4}. In addition, we have given the error bar plot for the mentioned
Pantheon datasets in Fig. \ref{Mu} along with our resulting model compared to the $%
\Lambda $CDM model (with $\Omega _{m}^{0}=0.3$, $\Omega _{\Lambda }^{0}=0.7$
and $H_{0}=69$ $km.s^{-1}.Mpc^{-1}$). The graph illustrates that our model
fits the observational Pantheon datasets well.

\begin{widetext}		

\begin{figure}[H]
\centering
\includegraphics[scale=0.55]{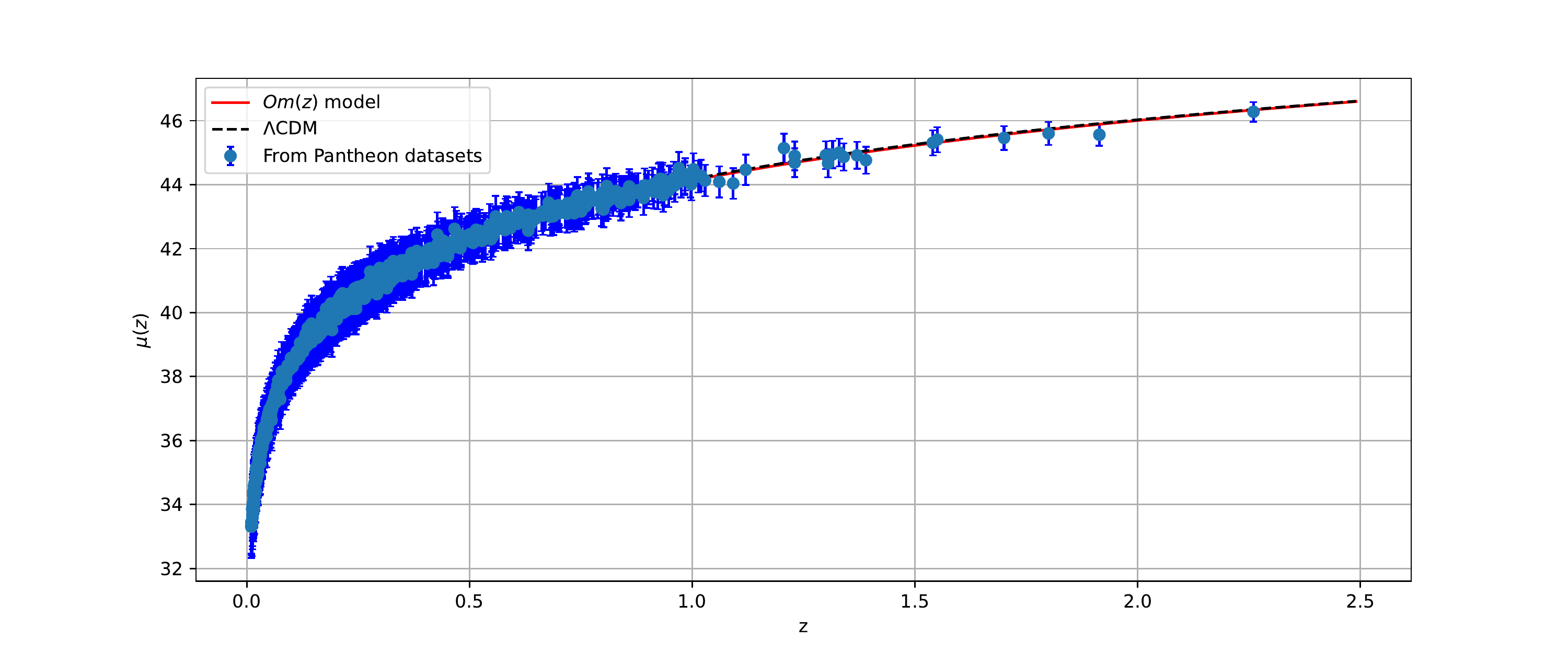}
\caption{The confidence curves at $1-\sigma$ and $2-\sigma$ and posterior distributions for the model parameters using Pantheon datasets. The dark pink shaded areas represent the $1-\sigma$ confidence level (CL), while the light pink shaded areas represent the $2-\sigma$ CL. The parameter constraint values are also presented at the $1-\sigma$ CL.}
\label{Mu}
\end{figure}

\begin{figure}[H]
\centering
\includegraphics[scale=0.75]{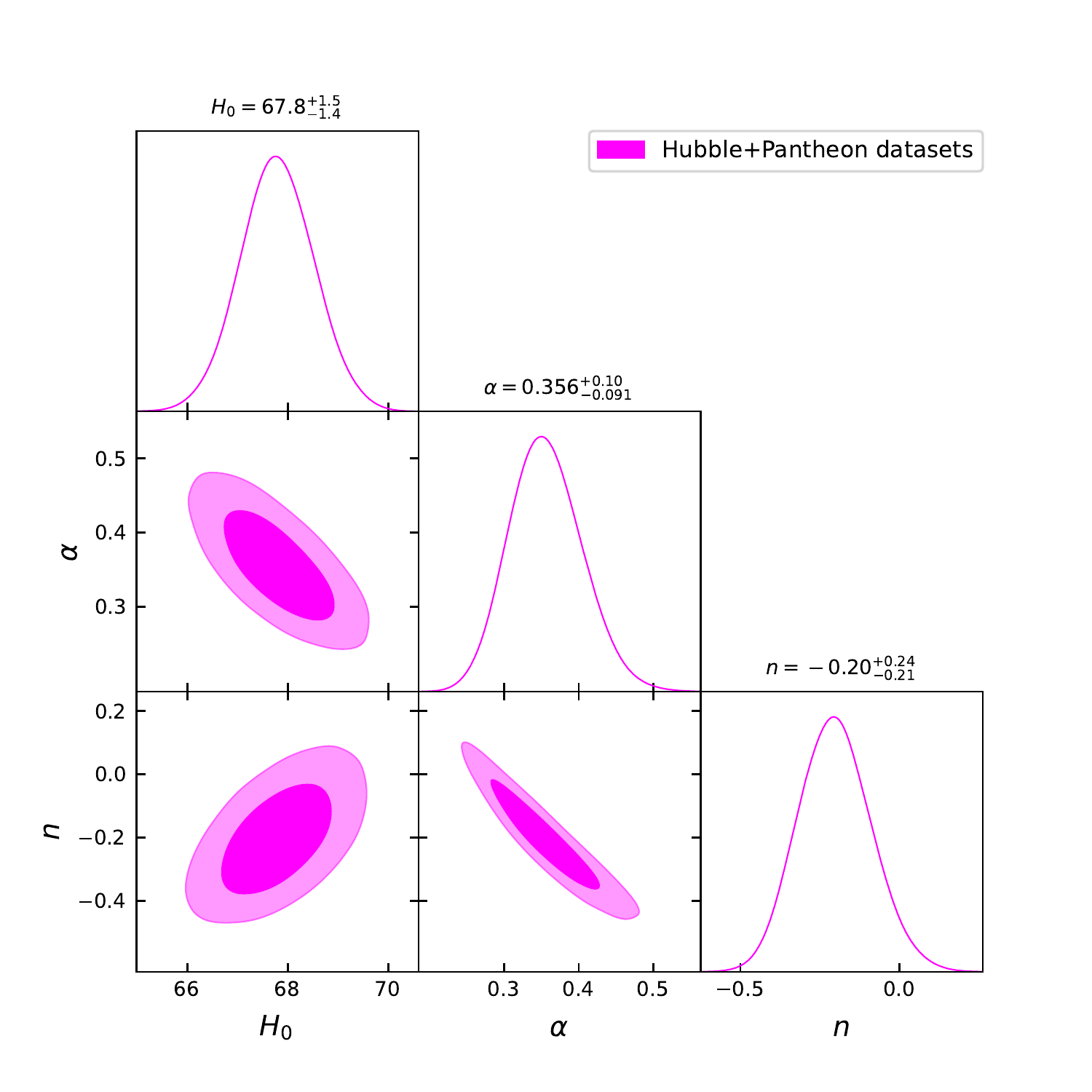}
\caption{The confidence curves at $1-\sigma$ and $2-\sigma$ and posterior distributions for the model parameters using Hubble+Pantheon datasets. The dark pink shaded areas represent the $1-\sigma$ confidence level (CL), while the light pink shaded areas represent the $2-\sigma$ CL. The parameter constraint values are also presented at the $1-\sigma$ CL.}
\label{H+SN}
\end{figure}	

\end{widetext}

\subsection{Baryon Acoustic Oscillations (BAO) datasets}

BAO are fluctuations in the density of the observable baryonic matter of the
Universe induced by acoustic density waves in the early Universe's
primordial plasma. As shown in Tab. \ref{tab3}, the BAO distance datasets,
which include the 6dFGS, SDSS, and WiggleZ surveys, contain BAO values at
six unique redshifts. Also, the characteristic scale of BAO is governed by
the sound horizon $r_{s}$ at the epoch of photon decoupling $z_{\ast }$,
which is determined by the following relation: \newline
\begin{equation}
r_{s}(z_{\ast })=\frac{c}{\sqrt{3}}\int_{0}^{\frac{1}{1+z_{\ast }}}\frac{da}{%
a^{2}H(a)\sqrt{1+(3\Omega _{b0}/4\Omega _{\gamma 0})a}},
\end{equation}%
where, $\Omega _{b0}$ and $\Omega _{\gamma 0}$ are the current density of
baryons and photons, respectively.

The BAO sound horizon scale is used to calculate the angular diameter
distance $d_{A}$ and the Hubble expansion rate $H\left( z\right) $ as a
function of redshift $z$. If the observed angular separation value of the
BAO feature is represented by $\triangle \theta $ in the two-point
correlation function of the galaxy distribution on the sky, and the observed
redshift separation value of the BAO feature is represented by $\Delta z$ in
the same two-point correlation function along the line of sight, we have the
relation,%
\begin{equation}
\triangle \theta =\frac{r_{s}}{d_{A}(z)},
\end{equation}%
where%
\begin{equation}
d_{A}(z)=c\int_{0}^{z}\frac{dz^{^{\prime }}}{H(z^{^{\prime }})},  \label{4d}
\end{equation}%
and%
\begin{equation}
\triangle z=H(z)r_{s}.
\end{equation}

In this study, we employed BAO datasets of 6 points for $d_{A}(z_{\ast
})/D_{V}(z_{BAO})$, which collected from the Refs. \cite{BAO1, BAO2, BAO3,
BAO4, BAO5, BAO6} and presented in Tab. \ref{tab3}, where $z_{\ast }\approx 1091$ is
the redshift at the epoch of photon decoupling and $d_{A}(z)$ is the
co-moving angular diameter distance combined with the dilation scale $%
D_{V}(z)=\left[ d_{A}(z)^{2}cz/H(z)\right] ^{1/3}$. In addition, it must be noted that the sound horizon and the redshift of decoupling depend on the baryon and radiation densities, which are not explicitly included in our parametrization. However, we assume that fixing the redshift of decoupling at $z_{\ast}\approx 1091$ is a reasonable model-independent approximation, as it is consistent with previous measurements and theoretical expectations. 

The chi-square function for the BAO datasets is defined as,

\begin{equation}
\chi _{BAO}^{2}=X^{T}C_{BAO}^{-1}X,
\end{equation}%
where 
\begin{widetext}
\begin{equation*}
X=\left( 
\begin{array}{c}
\frac{d_{A}(z_{\star })}{D_{V}(0.106)}-30.95 \\ 
\frac{d_{A}(z_{\star })}{D_{V}(0.2)}-17.55 \\ 
\frac{d_{A}(z_{\star })}{D_{V}(0.35)}-10.11 \\ 
\frac{d_{A}(z_{\star })}{D_{V}(0.44)}-8.44 \\ 
\frac{d_{A}(z_{\star })}{D_{V}(0.6)}-6.69 \\ 
\frac{d_{A}(z_{\star })}{D_{V}(0.73)}-5.45%
\end{array}%
\right) \,,
\end{equation*}%
and the inverse covariance matrix $C_{BAO}^{-1}$ is represented in \cite%
{BAO6} as,

\begin{equation*}
C_{BAO}^{-1}=\left( 
\begin{array}{cccccc}
0.48435 & -0.101383 & -0.164945 & -0.0305703 & -0.097874 & -0.106738 \\ 
-0.101383 & 3.2882 & -2.45497 & -0.0787898 & -0.252254 & -0.2751 \\ 
-0.164945 & -2.454987 & 9.55916 & -0.128187 & -0.410404 & -0.447574 \\ 
-0.0305703 & -0.0787898 & -0.128187 & 2.78728 & -2.75632 & 1.16437 \\ 
-0.097874 & -0.252254 & -0.410404 & -2.75632 & 14.9245 & -7.32441 \\ 
-0.106738 & -0.2751 & -0.447574 & 1.16437 & -7.32441 & 14.5022%
\end{array}%
\right) \,.
\end{equation*}	

\begin{table}[H]

\begin{center}

\begin{tabular}{ccccccc}
\hline\hline
$z_{BAO}$ & $0.106$ & $0.2$ & $0.35$ & $0.44$ & $0.6$ & $0.73$ \\ \hline
$\frac{d_{A}(z_{\ast })}{D_{V}(z_{BAO})}$ & $30.95\pm 1.46$ & $17.55\pm 0.60$
& $10.11\pm 0.37$ & $8.44\pm 0.67$ & $6.69\pm 0.33$ & $5.45\pm 0.31$ \\ 
\hline\hline
\end{tabular}
\caption{Values of $d_{A}(z_{\ast })/D_{V}(z_{BAO})$ for distinct values of $z_{BAO}$.}
\label{tab3}
\end{center}
\end{table}

\end{widetext}

By minimizing $\chi _{Hubble}^{2}+\chi _{Pan}^{2}+\chi _{BAO}^{2}$, the
constraints from the combination Hubble+Pantheon+BAO datasets are shown in
Fig. \ref{H+SN+BAO} and numerical findings presented in Tab. \ref{tab4}.

\begin{widetext}	

\begin{figure}[H]
\centering
\includegraphics[scale=0.75]{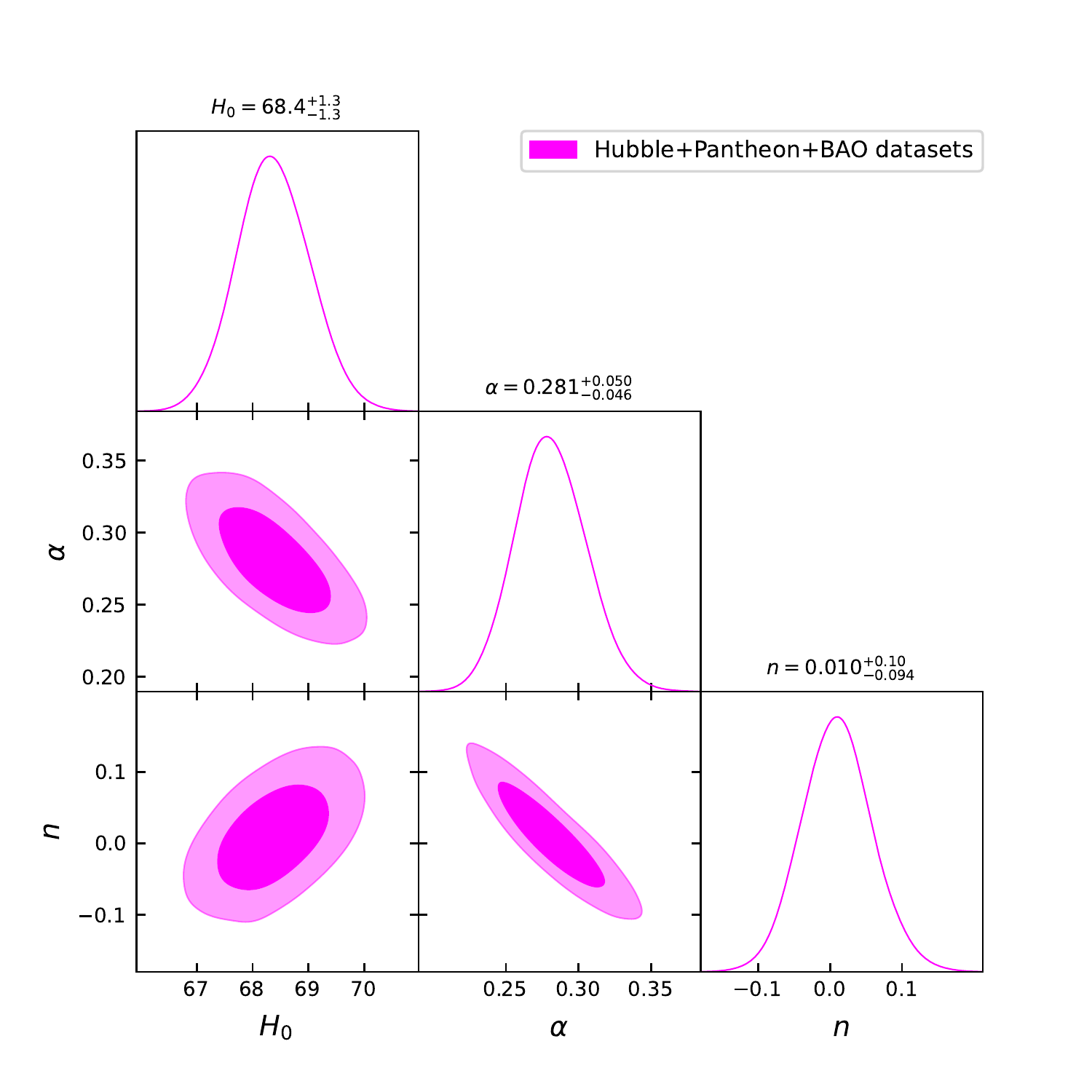}
\caption{The confidence curves at $1-\sigma$ and $2-\sigma$ and posterior distributions for the model parameters using Hubble+Pantheon+BAO datasets. The dark pink shaded areas represent the $1-\sigma$ confidence level (CL), while the light pink shaded areas represent the $2-\sigma$ CL. The parameter constraint values are also presented at the $1-\sigma$ CL.}
\label{H+SN+BAO}
\end{figure}

\begin{table*}[!htbp]
\begin{center}
\begin{tabular}{l c c c c c c}
\hline\hline 
$datasets$              & $H_{0}$ $(km/s/Mpc)$ & $\alpha$ & $n$ & $\omega_0$ & $q_{0}$ & $z_{tr}$ \\
\hline
$Priors$   & $(60,80)$ & $(0,1)$  & $(-10,10)$  & $-$ & $-$ & $-$\\
$Hubble$ & $67.8^{+1.7}_{-1.7}$ & $0.38^{+0.12}_{-0.12}$  & $-0.25^{+0.28}_{-0.25}$  & $-0.62^{+0.12}_{-0.12}$ & $-0.43^{+0.18}_{-0.18}$ & $0.710\pm 0.18$\\
$Hubble+Pantheon$   & $67.8^{+1.5}_{-1.4}$ & $0.356^{+0.10}_{-0.091}$  & $-0.20^{+0.24}_{-0.21}$  & $-0.644^{+0.1}_{-0.091}$ & $-0.466^{+0.15}_{-0.1365}$ & $0.732_{-0.17}^{+0.31}$\\
$Hubble+Pantheon+BAO$   & $68.4^{+1.3}_{-1.3}$ & $0.281^{+0.050}_{-0.046}$  & $0.010^{+0.10}_{-0.094}$  & $-0.719^{+0.05}_{-0.046}$ & $-0.5785^{+0.075}_{-0.069}$ & $0.701_{-0.15}^{+0.23}$\\

\hline\hline
\end{tabular}
\caption{A summary of the MCMC findings obtained from several datasets.}
\label{tab4}
\end{center}
\end{table*}

\end{widetext}

\section{Discussion of the findings}

\label{sec4}

\begin{figure}[]
\centering
\includegraphics[scale=0.70]{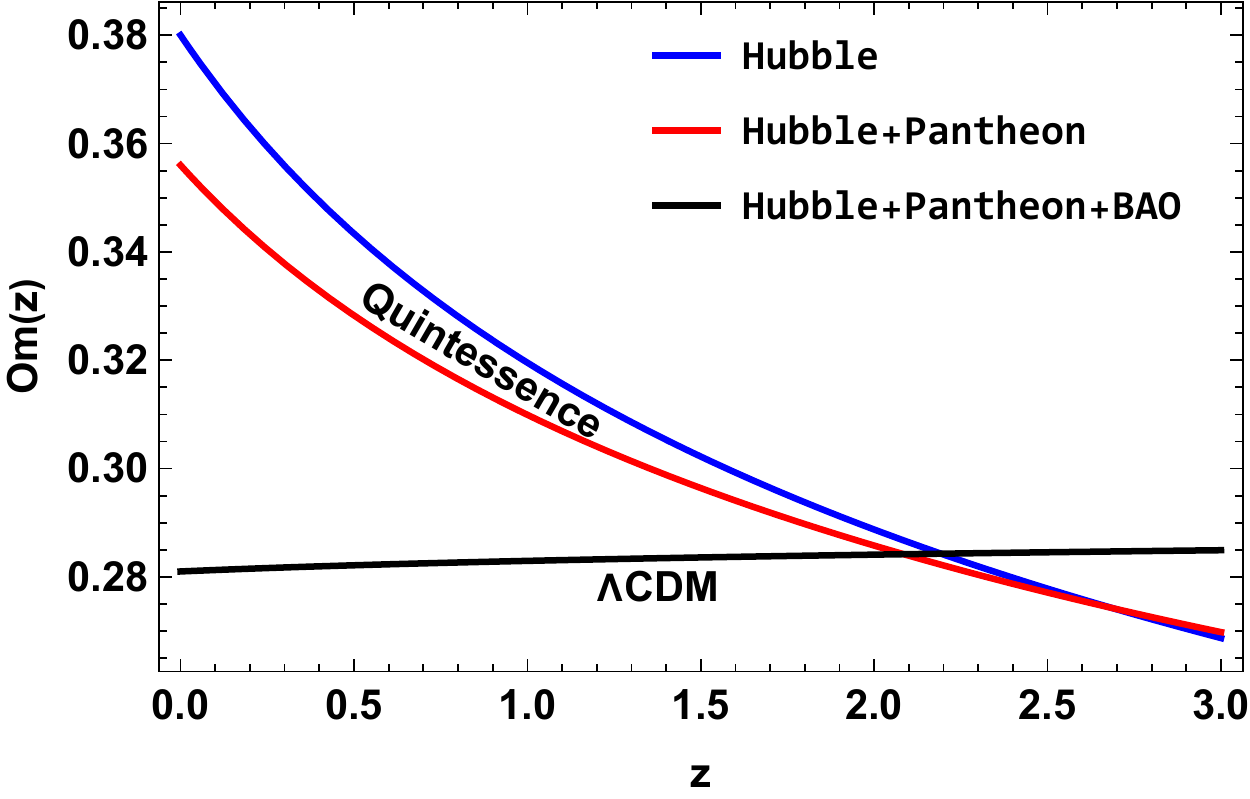}
\caption{The behavior of the $Om(z)$ diagnostic vs. redshift $z$.}
\label{Om}
\end{figure}

\begin{figure}[]
\centering
\includegraphics[scale=0.70]{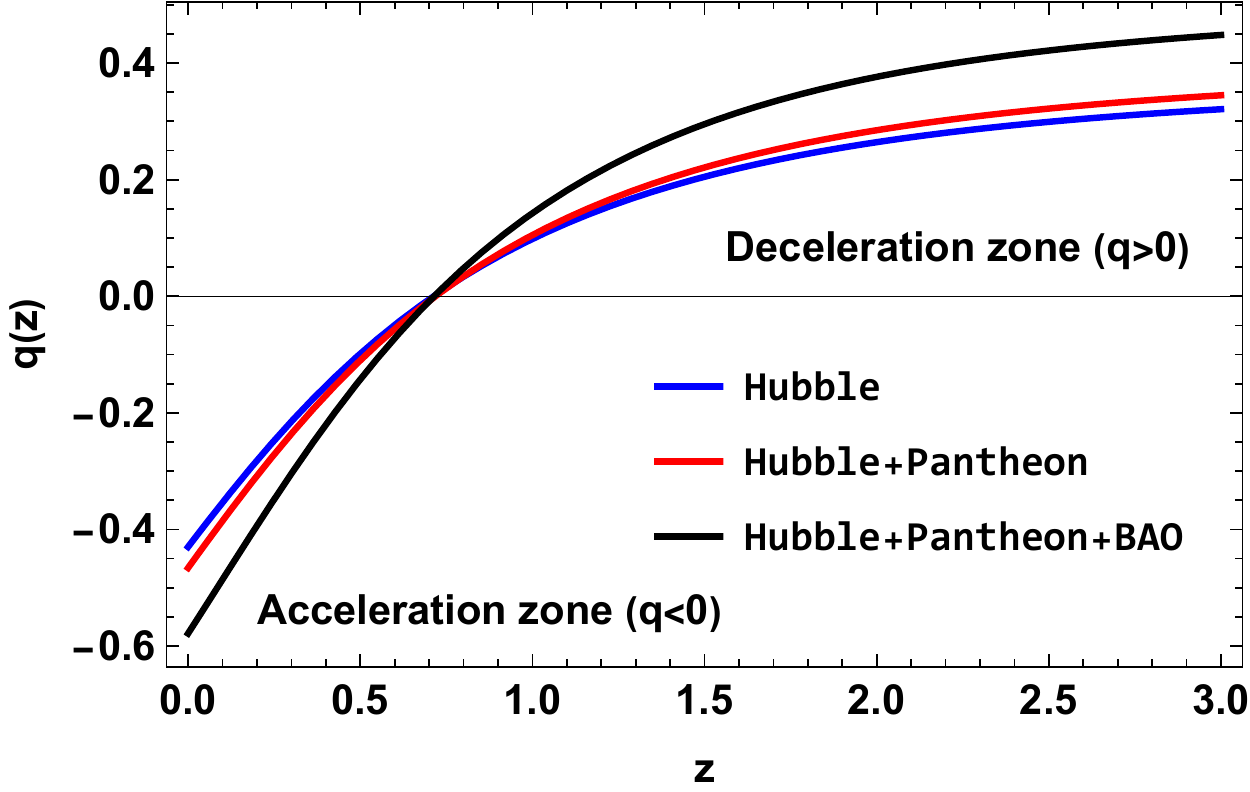}
\caption{The behavior of the deceleration parameter $q$ vs. redshift $z$.}
\label{q}
\end{figure}

\begin{figure}[]
\centering
\includegraphics[scale=0.70]{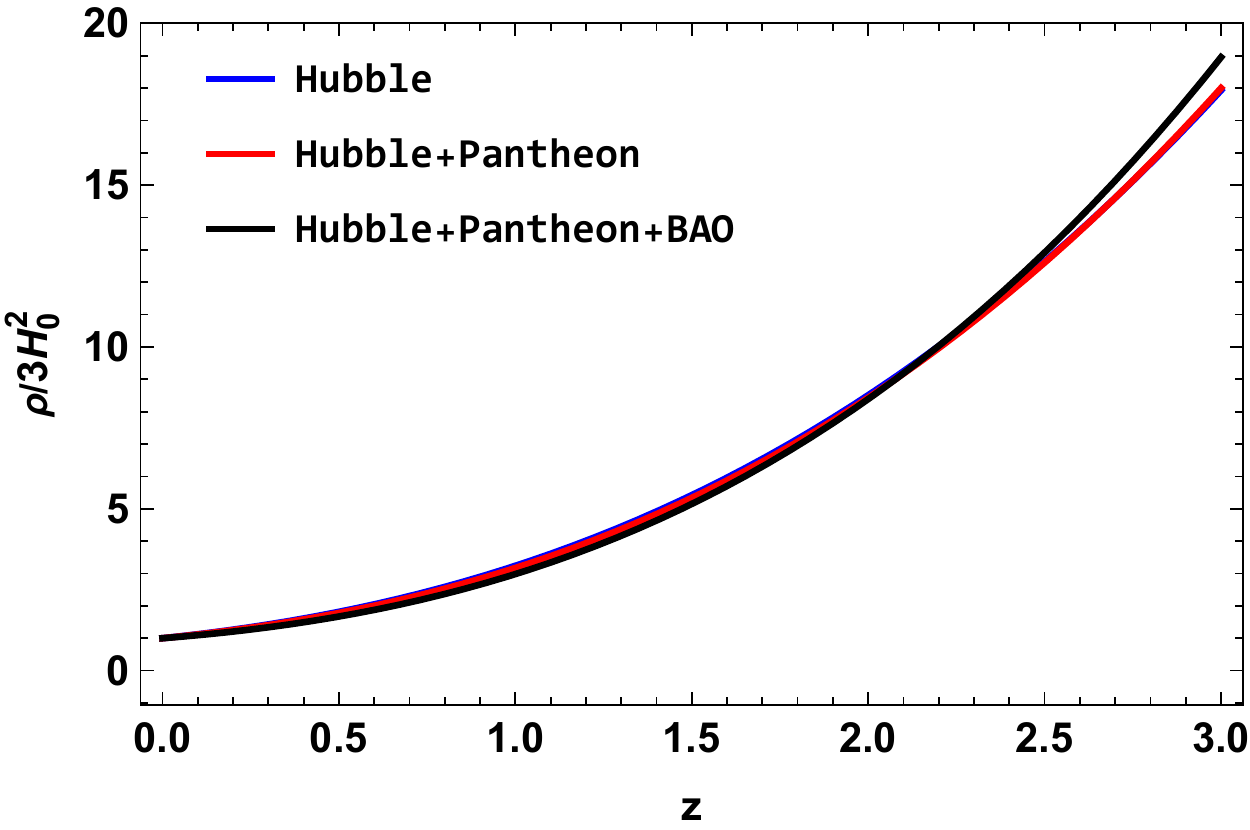}
\caption{The behavior of the density parameter $\rho$ vs. redshift $z$.}
\label{rho}
\end{figure}

\begin{figure}[]
\centering
\includegraphics[scale=0.70]{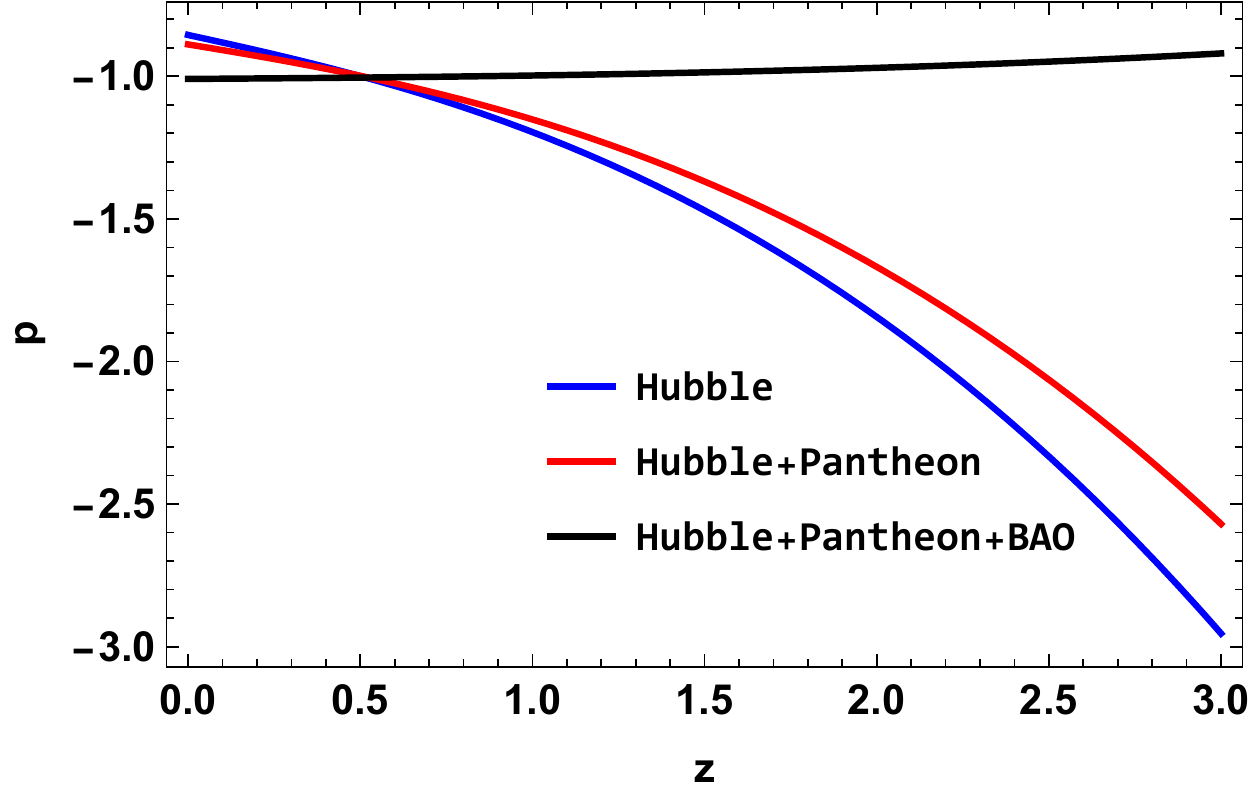}
\caption{The behavior of the pressure $p$ vs. redshift $z$.}
\label{P}
\end{figure}

\begin{figure}[]
\centering
\includegraphics[scale=0.70]{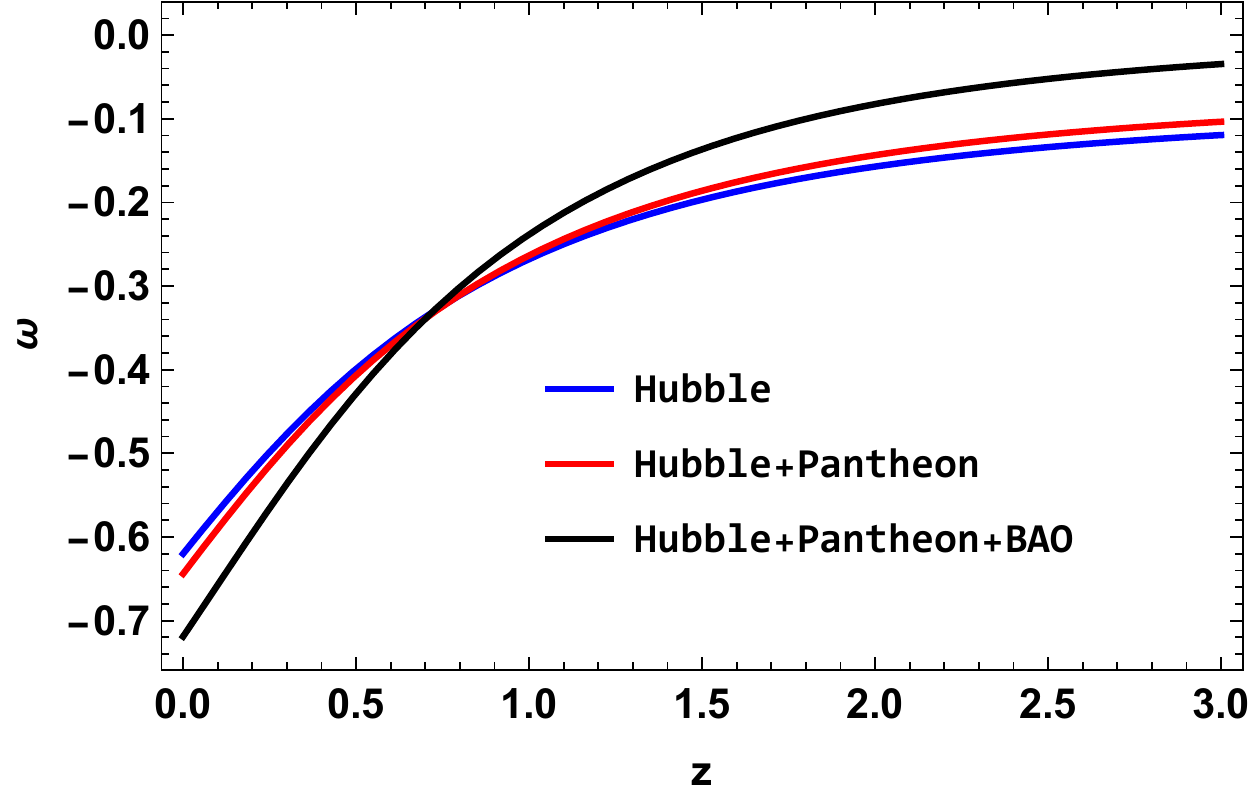}
\caption{The behavior of the EoS parameter $\omega$ vs. redshift $z$.}
\label{EoSS}
\end{figure}

In this section, we will discuss the findings of the statistical analysis
and their application to the previous cosmological parameters. The
investigation of cosmological parameters is an essential technique to
describe many characteristics of the Universe. The parameterizations of
various functions, plus some simple constants, are utilized to explain the
characteristics of cosmological parameters. These parameters, including the
expansion rate and curvature, describe the global dynamics of the Universe.
Here, we investigated several of the fundamental parameters of our current $Om(z)$
parameterization in FLRW Universe, such as the deceleration parameter, the
density parameter, the pressure, and the EoS parameter.

Initially, we examined various data samples and estimated the constraint
values for the model parameters $\alpha $, $n$ and $H_{0}$. We also
constructed two-dimensional likelihood contours with $1-\sigma $ and $%
2-\sigma $ errors and $68\%$ and $95\%$ CL for Hubble,
Hubble+Pantheon, and Hubble+Pantheon+BAO datasets (Figs. \ref{H}, \ref{H+SN}, and \ref{H+SN+BAO} show
this). The likelihood functions for all datasets are extremely well fitted
to a Gaussian distribution function, as shown in Figs. \ref{H}, \ref{H+SN}, and \ref{H+SN+BAO}. At first, we
examined the Hubble datasets, which contain 31 data points. Thus, we got
the value: $0.38_{-0.12}^{+0.12}$ for the model parameter $\alpha $, and
the constraint value is $-0.25_{-0.25}^{+0.28}$ for the parameter $n$, which
differentiates between different DE models. The value of the parameter $n$
from the Hubble datasets shows that $Om(z)$ has a negative slope indicating
the quintessence epoch. For combined
Hubble+Pantheon datasets, we obtain the values, $\alpha
=0.356_{-0.091}^{+0.10}$ and $n=-0.20_{-0.21}^{+0.24}$, which indicates the
same behavior as the Hubble datasets. Finally, we get these values from the
combined Hubble+Pantheon+BAO datasets: $\alpha =0.281_{-0.046}^{+0.050}$
and $n=0.010_{-0.094}^{+0.10}$, which approximately corresponds to the
constant slope i.e. the cosmological constant. The best-fit curves of $Om(z)$ diagnostic with various values of model parameters constrained from the
Hubble, Hubble+Pantheon, and Hubble+Pantheon+BAO datasets is
shown in Fig. \ref{Om} with further details. It is important to note that Fig. \ref{Om} shows only the best-fit model for each dataset, and that it does not necessarily represent the full range of allowed variations in $Om(z)$. Indeed, as shown in Tab. \ref{tab4}, all datasets are compatible with a constant $Om(z)$, corresponding to the standard $\Lambda$CDM model, within the uncertainties.

In addition, to compare our $Om(z)$ parameterization to the $\Lambda $CDM
model, we examined the Hubble parameter $H(z)$ curve and distance modulus $%
\mu \left( z\right) $ curve with the constraint values of model parameters $%
\alpha $ and $n$ for Hubble and Pantheon samples datasets, as shown in
Figs. \ref{Hz} and \ref{Mu}. The red line in the graphics indicates the theoretical curve for the best-fit values obtained by the Hubble and Pantheon datasets. It is noticed that our $Om(z)$ parameterization matches the
observational results well in both cases. Furthermore, it can be shown that
our parameterization is pretty similar to the curve of the $\Lambda $CDM
model (the black
dashed line). Here, we estimated the current Hubble parameter values $\left(
z=0\right) $ to be: $H_{0}=67.8_{-1.7}^{+1.7}$ $km/s/Mpc$, $%
H_{0}=67.8_{-1.4}^{+1.5}$ $km/s/Mpc$, and $H_{0}=68.4_{-1.3}^{+1.3}$ $km/s/Mpc$ for the Hubble, Hubble+Pantheon, and Hubble+Pantheon+BAO datasets, respectively,
which are very consistent with recent Planck's measurements \cite{Planck2020} and other studies in a similar context \cite{Chen1, Chen2, Aubourg, Capozziello}.

Fig. \ref{q} depicts the best-fit curve of $q(z)$ for each datasets to show the differences in the behavior of $q(z)$ for each dataset. Using
our $Om(z)$ parameterization, the current value of the deceleration
parameter (i.e. $z=0$) is approximated as $q_{0}=-0.43\pm 0.18$, $q_{0}=-0.466_{-0.1365}^{+0.15}$, and $%
q_{0}=-0.5785_{-0.069}^{+0.075}$ for the Hubble, Hubble+Pantheon,
and Hubble+Pantheon+BAO datasets, respectively. It is important to note that the values of $q_{0}$ constrained in this study are compatible with the value obtained in Refs. \cite{Capozziello, Mamon, Basilakos}. As a consequence, the
suggested model's results are consistent with current data \cite{Planck2020}. Furthermore, we
can see that the early Universe was in a decelerated period ($q>0$) of
expansion while the present Universe accelerated ($q<0$). Thus, the Universe
with our $Om(z)$ parameterization reflects a transition (i.e. $q=0$) with
signature flipping at $z_{tr}=0.710\pm 0.18$, $z_{tr}=0.732_{-0.17}^{+0.31}$, and $%
z_{tr}=0.701_{-0.15}^{+0.23}$ for the Hubble, Hubble+Pantheon, and
Hubble+Pantheon+BAO datasets, respectively. These transition redshift
estimates are consistent with the previously constrained value of \cite{Farooq}, $%
z_{tr}=0.72$. The transition from deceleration to acceleration in the $Om(z)
$ parameterization process occurs at a redshift of $z_{tr}=0.701$ in the combined Hubble+Pantheon+BAO datasets, which is
consistent with the results of \cite{Farooq, Jesus, Garza} As a result, we see that our model
supports the most current scientific findings in all three scenarios.

Fig. \ref{rho} depicts the predicted positive behavior of the density parameter as
it decreases with the expansion of the Universe in the current time. However, we would like to clarify that the density parameter being referred to in this figure corresponds to the total matter-energy density of the Universe, which includes both dark matter and DE. Therefore, this density parameter should increase with redshift for any type of DE model, including the cosmological constant, quintessence, or phantom models.
 Fig. \ref{P}
displays the negative behavior of the pressure $p$, reflecting the Universe's
late-time cosmic acceleration. It can be shown that the Hubble,
Hubble+Pantheon, and Hubble+Pantheon+BAO datasets display similar pressure
evolutions in the past. However, the current negative behavior indicates
acceleration. Furthermore, it is generally understood that the EoS parameter
also plays an important role in explaining the many energy-dominated
evolution processes of the Universe. The current state of the Universe may
be predicted via the quintessence phase ($-1<\omega_{DE} <-\frac{1}{3}$) or the
phantom phase ($\omega_{DE} <-1$). Fig. \ref{EoSS} depicts the best-fit curve of $\omega
(z)$. So, with the current model, we got $\omega _{0}=-0.62\pm 0.12$, $\omega _{0}=-0.644_{-0.091}^{+0.1}$,
and $\omega _{0}=-0.719_{-0.046}^{+0.05}$ for the Hubble,
Hubble+Pantheon, and Hubble+Pantheon+BAO datasets, respectively. We note that the quintessence-like behavior of the EoS parameter for our $Om(z)$ parameterization, as seen in Fig. \ref{EoSS}, is expected due to the dominant pressureless matter contribution at high redshifts. Our findings on $\omega (z)$ are
consistent with the findings of certain observational studies \cite{Hernandez, Zhang}. The current
values for various cosmological parameters $H_{0}$, $q_{0}$, $z_{tr}$ and $%
\omega _{0}$ are summarized in Tab. \ref{tab4}.

\section{Final Remarks and Perspectives}
\label{sec5}

The $Om(z)$ diagnostic method holds significant importance in testing cosmology within the framework of GR as well as various modified theories of gravity. In essence, this research paper presents a novel approach to parameterizing the $Om(z)$ diagnostic and examines its behavior within the context of GR. This diagnostic is specifically designed to make predictions for both phantom and quintessence models of Dark Energy, and the model parameters are meticulously determined by analyzing observational data, including 31 data points from Hubble expansion observations, six Baryon Acoustic Oscillation (BAO) data points, and an extensive datasets of 1048 SNe Ia from Pantheon. To further refine the cosmological model, the Markov Chain Monte Carlo (MCMC) approach has been employed. 

Our investigation shows that the new parametrization of $Om(z)$ stands in good agreement with the Hubble expansion observations. The best-fit values of the model with Hubble data are: $H_0 = 67.8^{+1.7}_{-1.7}$ $km/s/Mpc$, $\alpha = 0.38^{+0.12}_{-0.12}$ and $n = - 0.25^{+0.28}_{-0.25}$. In the next phase, we considered both Hubble+ Pantheon datasets and obtain best-fit values: $H_0 = 67.8^{+1.5}_{-1.4}$ $km/s/Mpc$, $\alpha = 0.356^{+0.10}_{-0.091}$ and $n = - 0.20^{+0.24}_{-0.21}$, which is comparatively well constrained in comparison to the previous results obtained by us.
Further, to enhance the results, we consider Hubble+Pantheon+ BAO and obtain: $\alpha=0.281^{+0.050}_{-0.046}$, $n=0.010^{+0.10}_{-0.094}$ and $H_0 = 68.4^{+1.3}_{-1.3}$ $km/s/Mpc$.

These results provide valuable insights into the evolution of the cosmos and enhance our understanding of the nature of Dark Energy. Furthermore, the proposed parameterization of the $Om(z)$ diagnostic capable of explaining phantom and quintessence has the potential to facilitate the testing of alternative Dark Energy models, thereby leading to a better understanding of the Universe's evolution.

This novel parameterization of the $Om(z)$ diagnostic can be used in different modified theories of gravity, including $f(R)$ gravity, $f(Q)$ gravity, Rastall gravity etc., to examine the behavior of Universe's evolution and other cosmographic parameters as well. We keep this as a future prospect of the study.

\section*{Acknowledgments}
This research is funded by the Science Committee of the Ministry of Science and Higher Education of the Republic of Kazakhstan (Grant No. AP09058240). M. Koussour is thankful to Dr. Shibesh Kumar Jas Pacif, Centre for Cosmology and Science Popularization, SGT University for some useful discussions. D. J. Gogoi is thankful to Prof. U. D. Goswami, Dibrugarh University for some useful discussions.

\textbf{Data availability} All data used in this study are cited in the references and were obtained from publicly available sources.

\textbf{Conflict of interest} The authors declare that they have no known competing financial interests or personal relationships that could have appeared to influence the work reported in this paper.\newline

\end{document}